\documentstyle[prl,aps,twocolumn,epsfig]{revtex}

\newcommand{\IP}{\mbox{I}\!\mbox{P}}

\newcommand{\GeV}{{\rm\,GeV}}

\begin{document}

\title{Electroweak and finite width corrections to top quark decays into
  transverse and longitudinal $ W $-bosons}

\author{H.S.~Do, S.~Groote, J.G.~K\"orner and M.C.~Mauser}
\address{Institut f\"ur Physik der Johannes-Gutenberg-Universit\"at,
  Staudinger Weg 7, 55099 Mainz, Germany}

\maketitle

\begin{abstract}
 We calculate the electroweak and finite width  corrections to the decay of
 an unpolarized top quark into a bottom quark and a $ W $-gauge boson where
 the helicities of the $ W $ are specified as longitudinal, transverse-plus
 and transverse-minus. Together with the $ O(\alpha_s) $ corrections these
 corrections may become relevant for the determination of the mass of the
 top quark through angular decay measurements.
\end{abstract}


\section{Introduction} 
 Because the Kobayashi-Maskawa matrix element $V_{tb}$ is very close to 1 the
 dominating decay mode of the top quark is into the channel
 $ t \rightarrow X_{b} + W^{+} $. The helicity content of the $ W $-boson
 (transverse-plus, transverse-minus and longitudinal) in this decay can be
 probed through a measurement of the shape of the lepton spectrum in the decay
 of the $ W $-boson as was recently done by the CDF Collaboration~\cite{cdf}.
 Of particular interest is the size of the longitudinal contribution which
 encodes the physics of the spontaneous symmetry breaking of electroweak
 symmetry. The transverse-plus contribution vanishes at the Born term level.
 Any significant deviation from this value would point to sizeable radiative
 corrections or a non-SM $ (V+A) $ admixture in the weak $ t \rightarrow b $
 current transition.

 A first measurement of the helicity content of the $ W $-boson was recently
 carried through by the CDF Collaboration~\cite{cdf}. Their result is
 \begin{equation} \label{CDFlong}
   \Gamma_{L}/\Gamma = 0.91 \pm 0.37(stat) \pm 0.13(syst) \qquad
 \end{equation}
 where $ \Gamma_L $ denotes the rates into the longitudinal polarization
 state of the $ W $-boson and $ \Gamma $ is the total rate. The CDF
 Collaboration has also quoted a value for the transverse-plus contribution
 $ \Gamma_{+}/\Gamma $ which was obtained by fixing the longitudinal
 contribution to its Standard Model (SM) value $ \Gamma_{L}/\Gamma = 0.7 $.
 They obtained
 \begin{equation} \label{CDFtrans+}
   \Gamma_{+}/\Gamma = 0.11 \pm 0.15.
 \end{equation}
 The measured values of the two normalized partial helicity rates are well
 within SM expectations which, at the Born term level, are
 $ \Gamma_{L}/\Gamma \approx 0.70 $ and $ \Gamma_{+}/\Gamma = 0$. However, the
 errors on this measurement are too large to allow for a significant test of
 the SM predictions. The errors will be much reduced when larger data samples
 become available in the future from TEVATRON RUN II, and, at a later stage,
 from the LHC. Optimistically the measurement errors can eventually be reduced
 to the $ (1-2)\% $ level~\cite{willenbrock}. If such a level of accuracy can
 in fact be reached it is important to take into account the radiative
 and finite width corrections to the different helicity rates.
 
 The radiative corrections to the top width are rather large. Relative to the
 $ m_b = 0 $ Born term rate they amount to $ -8.54\% $ (QCD one-loop)~%
 \cite{c14,c15,c16,c17,c18,ghinculov}, $ +1.54\% $ (electroweak one-loop)~%
 \cite{c14,EMMS91,bruecher} and $ -2.05\% $ ($ -2.16\% $) (QCD two-loop;
 approximate)~\cite{CM99} (\hspace{-2mm} \cite{CHSS99}).
 The $ m_b \neq 0 $ and finite width corrections reduce the Born width by
 $ 0.27\% $~\cite{FGKLM,FGKM1,FGKM2} (for $ m_b = 4.8 \GeV $~\cite{pivovarov})
 and $ 1.55\% $~\cite{jk93}, respectively. Given the fact that the radiative and
 finite width corrections to the total rate are sizeable it is important to
 know also the respective corrections to the partial longitudinal and transverse
 rates. The QCD one-loop corrections to the partial helicity rates were given
 in~\cite{FGKLM,FGKM1,FGKM2}. In this letter we provide the missing one-loop
 electroweak (EW) corrections and the finite width (FW) corrections to the
 partial helicity rates. Note that to leading order the finite width
 correction is also a one-loop effect. 

 The angular decay distribution for the decay process
 $ t \rightarrow X_b + W^{+} $ followed by
 $ W^{+} \rightarrow l^{+} + \nu_{l} $ (or by
 $ W^{+} \rightarrow \bar{q} + q $) is given by (see e.g.\ \cite{FGKM1})
 \begin{eqnarray}\label{ang}
   \frac{d \Gamma}{d \cos \theta} & = &
   \frac{3}{8} (1 + \cos \theta)^2 \Gamma_{+}
 + \frac{3}{8} (1 - \cos \theta)^2 \Gamma_{-} \nonumber \\ & & \qquad
 + \frac{3}{4} \sin^2 \theta\ \Gamma_{L}.
 \end{eqnarray}
 where $ \Gamma_{+} $, $ \Gamma_{-} $ and $ \Gamma_{L} $ denote the partial
 rates into transverse-plus, transverse-minus and longitudinal $ W $-bosons.
 Integrating over $ \cos \theta $ one recovers the total rate
 \begin{equation}
   \Gamma = \Gamma_{+} + \Gamma_{-} + \Gamma_{L}.
 \end{equation}

 One can also define a forward-backward asymmetry by considering the rate
 in the forward hemisphere $ \Gamma_F $ and in the backward hemisphere
 $ \Gamma_B $. The forward-backward asymmetry $ A_{FB} $ is then given by
 \begin{equation}
 \label{forward-backward}
   A_{FB} = \frac{\Gamma_F - \Gamma_B}{\Gamma_F + \Gamma_B} =
   \frac{3}{4} \, \frac{\Gamma_{+} - \Gamma_{-}}
   {\Gamma_{+} + \Gamma_{-} + \Gamma_{L}}.
 \end{equation}

 The angular decay distribution is described in cascade fashion, i.e.\ the polar
 angle $ \theta $ is measured in the $ W $ rest frame where the lepton pair
 or the quark pair emerges back-to-back. The angle $ \theta $ denotes the polar
 angle between the $ W^{+} $ momentum direction and the antilepton $ l^{+} $
 (or the antiquark $ \bar{q} $). The various contributions in (\ref{ang}) are
 reflected in the shape of the lepton energy spectrum in the rest frame of the
 top quark. From the angular factors in (\ref{ang}) it is clear that the
 contribution of $ \Gamma_{+} $ makes the lepton spectrum harder while
 $ \Gamma_{-} $ softens the spectrum where the hardness or softness is gauged
 relative to the longitudinal contribution. The only surviving contribution in
 the forward direction $ \theta = 0 $ comes from $ \Gamma_{+} $. The fact that
 $ \Gamma_{+} $ is predicted to be quite small implies that the lepton spectrum
 will be soft. The CDF measurement of the helicity content of the $ W^{+} $ in
 top decays was in fact done by fitting the values of the helicity rates to the
 shape of the lepton's energy spectrum.
 \begin{figure}
  \noindent
  \psfig{file=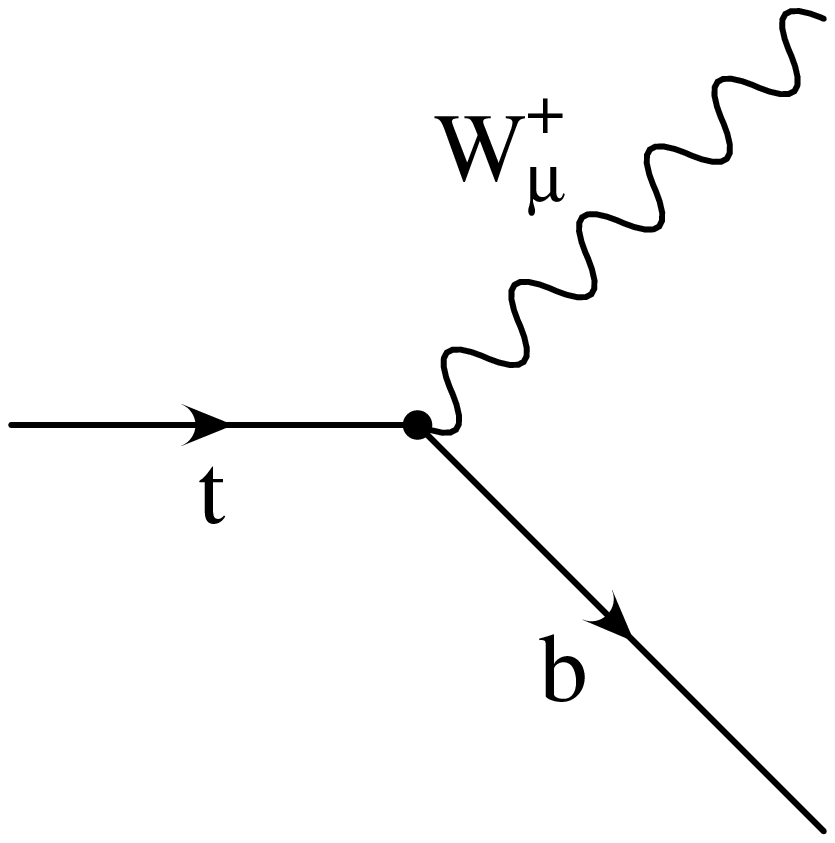,width=2.7cm}
  \psfig{file=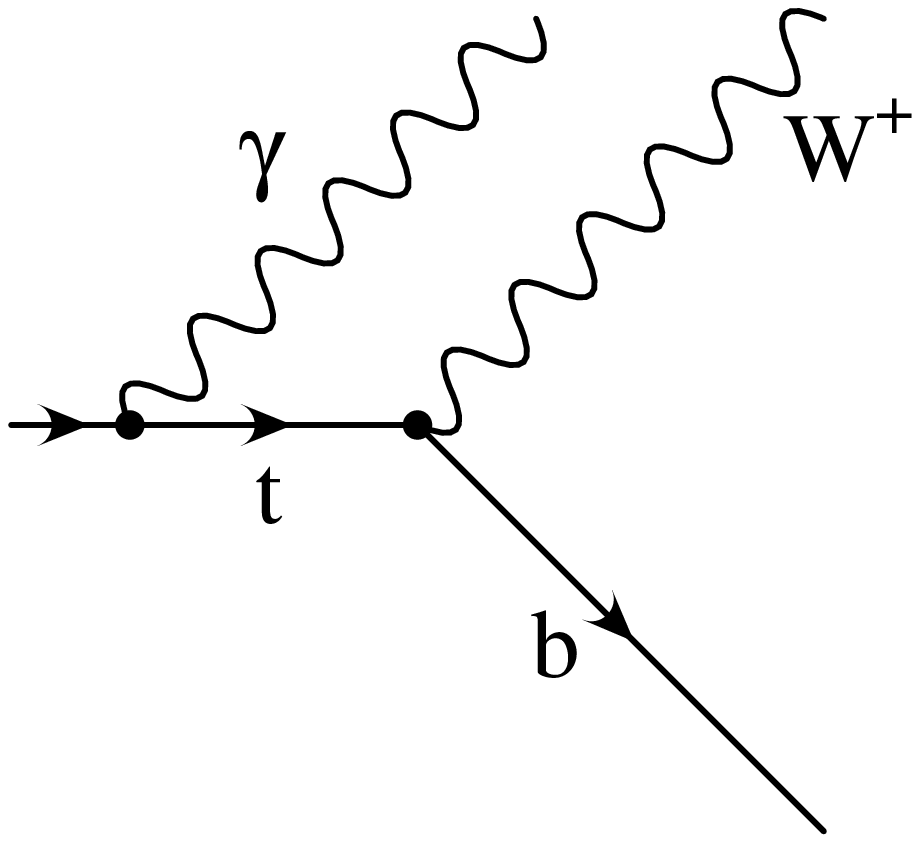,width=2.7cm}
  \psfig{file=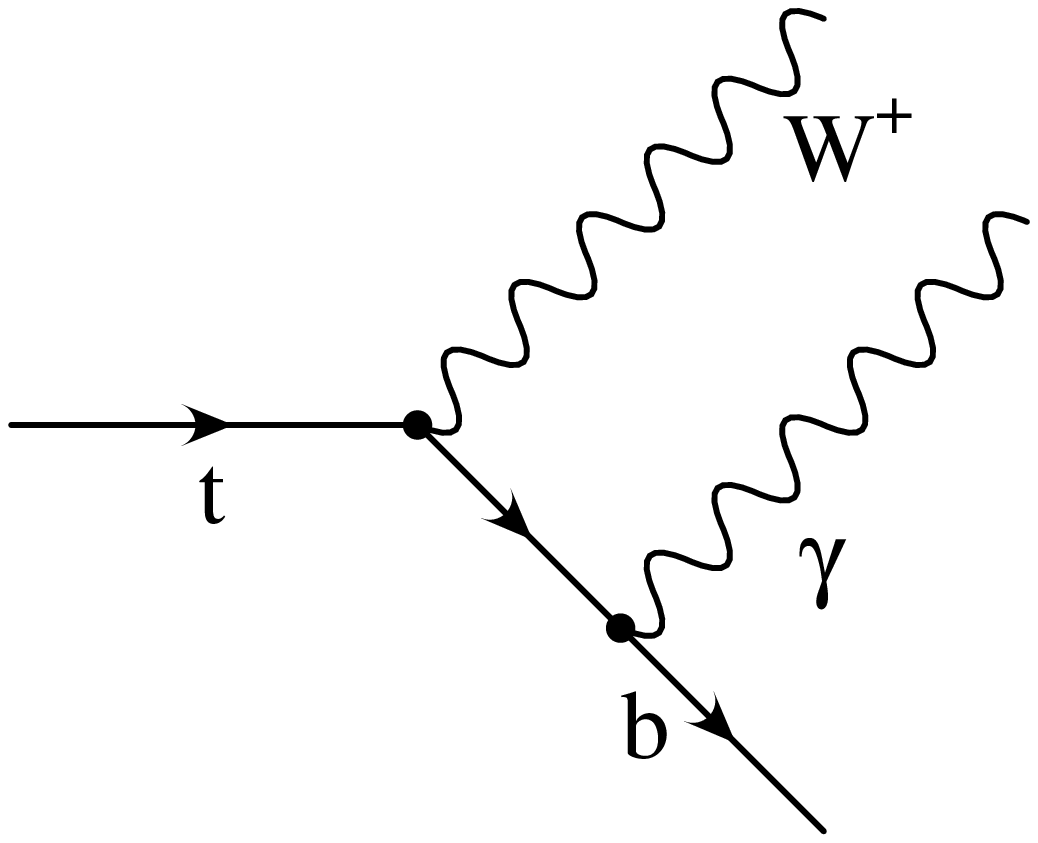,width=2.7cm} \newline
  \psfig{file=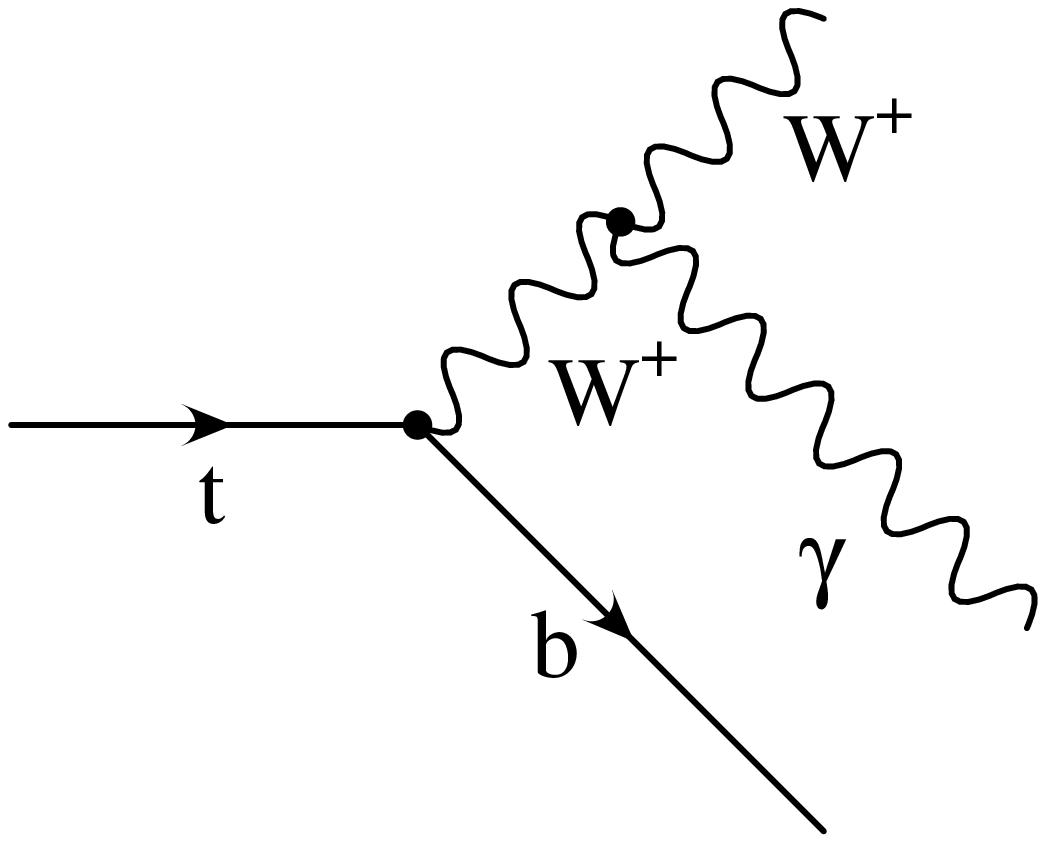,width=2.7cm}
  \psfig{file=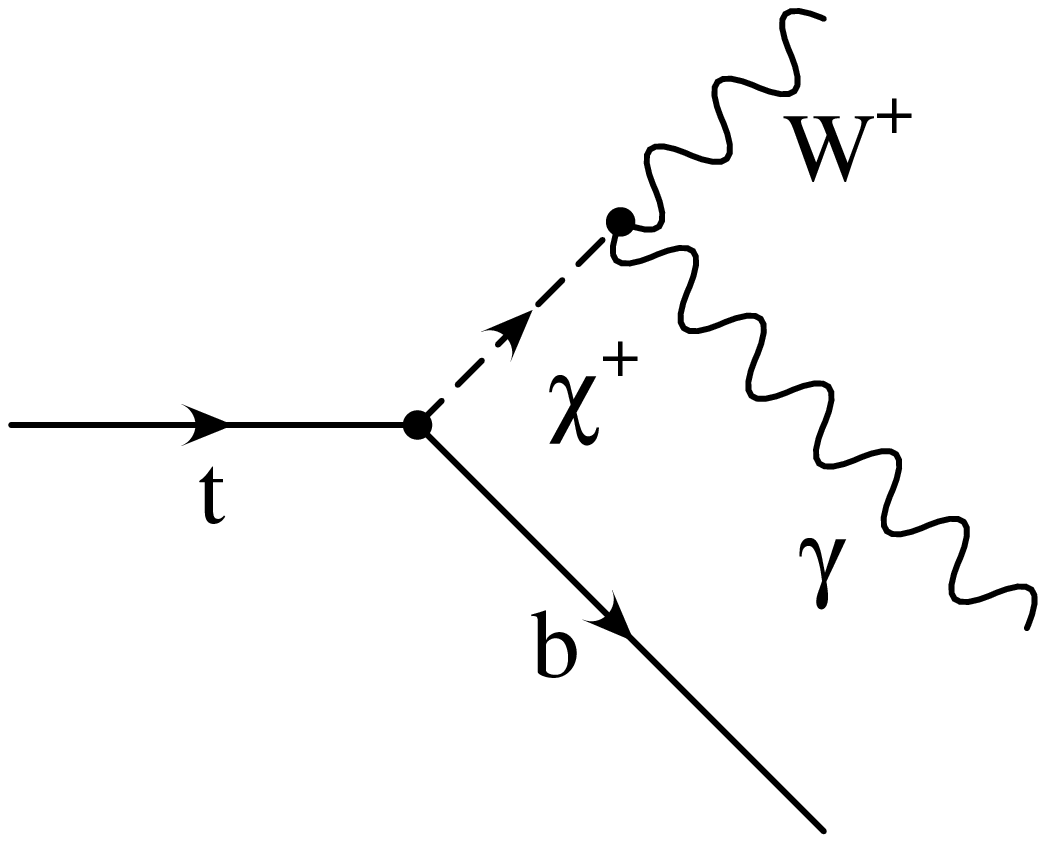,width=2.7cm}
  \caption{Born and electroweak tree-graph contributions to
  $ t \rightarrow b + W^+\, (\gamma) $. $ \chi^+ $ denotes the charged
  Goldstone boson.}
 \end{figure}
 The Born term contributions to the normalized helicity rates are given by
 $ \Gamma_{+}/\Gamma_{0} = 0 $,
 $ \Gamma_{-}/\Gamma_{0} = (2 x^2)/(1 + 2x^2) \, (= 0.297) $ and
 $ \Gamma_{L}/\Gamma_{0} = 1/(1+2x^2) \, (= 0.703) $, where $ \Gamma_{0} $ is
 the total Born term rate and $ x = m_W/m_t $ (see e.g.\ \cite{FGKM1}).
 Numerical values for the normalized helicity rates have been added in
 paranthesis using $ m_t = 175 \GeV $ and $ m_W = 80.419 \GeV $.

 We begin with the electroweak one-loop corrections. They consist of the four 
 tree diagram contributions shown in FIG.~1 and the one-loop contributions
 which are too numerous to be depicted in this letter.
 In the Feynman--'t Hooft gauge one has to calculate 18 different massive one-%
 loop three-point functions as well as the many massive one-loop two-point
 functions needed in the renormalization program. We have recalculated all one-%
 loop contributions analytically and have checked them numerically with the help
 of a XLOOPS/GiNaC package that automatically calculates one-loop three point
 functions \cite{bd02}. Our one-loop results agree with the results of
 \cite{c14}. The results are too lengthy to be reproduced here in analytical
 form. Analytical results will be given in a forthcoming publication \cite{DGKM}.

 The tree-graph contributions
 $ t \rightarrow b + W^{+} + \gamma $
 are determined in terms of the current transition matrix element
 $ M^{\mu} = \langle b,\gamma | J^{\mu} | t \rangle $.
 Upon squaring the current transition matrix element
 one obtains the hadron tensor
 $ H^{\mu \nu} = M^{\mu} M^{\nu \dagger} $.
 We have calculated the hadron tensor
 in the Feynman--'t Hooft gauge and obtain
 \begin{eqnarray} \label{hadrontensor}
   \lefteqn{H^{\mu \nu} =
   e^2 \frac{(p_t \cdot k)(p_b \cdot k)}{(q \cdot k)^2}
   \Big( \frac{Q_t}{p_t \cdot k} -
   \frac{Q_b}{p_b \cdot k}\Big)^2 \times} \nonumber \\ & &
   \Bigg\{ - \frac{p_t \cdot k}{p_b \cdot k}
   \bigg[ m_b^2 \Big( k^\mu p_t^\nu + k^\nu p_t^\mu
 - k \cdot p_t g^{\mu\nu} \Big) \nonumber \\ & &
 + i \Big( \epsilon^{\alpha \beta \mu \nu} p_b \cdot p_t
 - \epsilon^{\alpha \beta \gamma \nu} p_b^\mu p_{t,\gamma}
 + \epsilon^{\alpha \beta \gamma \mu} p_b^\nu p_{t,\gamma} \Big)
   k_\alpha p_{b,\beta} \nonumber \\ & &
 - i \Big( \epsilon^{\alpha \beta \mu \nu} k \cdot p_t
 - \epsilon^{\alpha \beta \gamma \nu} k^\mu p_{t,\gamma}
 + \epsilon^{\alpha \beta \gamma \mu} k^\nu p_{t,\gamma} \Big)
   k_\alpha p_{b,\beta} \bigg] \nonumber \\ & &
 + \frac{p_b \cdot k}{p_t \cdot k}
   \bigg[ m_t^2 \Big( k^\mu p_b^\nu + k^\nu p_b^\mu
 - k \cdot p_b g^{\mu\nu} -i \epsilon^{\alpha \beta \mu \nu}
   k_\alpha p_{b,\beta} \Big) \nonumber \\ & &
 - (p_t \cdot k) \Big( p_t^\mu p_b^\nu + p_t^\nu p_b^\mu
 - p_t \cdot p_b g^{\mu \nu} - i\epsilon^{\alpha \beta \mu \nu}
   p_{t,\alpha} p_{b,\beta} \Big) \nonumber \\ & &
 + (p_t \cdot k) \Big( k^\mu p_b^\nu + k^\nu p_b^\mu
 - k \cdot p_b g^{\mu\nu} - i \epsilon^{\alpha \beta \mu \nu}
   k_\alpha p_{b,\beta} \Big) \bigg] \nonumber \\ & &
 - (p_t \cdot p_b) \Big(k^\mu p_b^\nu + k^\nu p_b^\mu - 
   k \cdot p_b g^{\mu\nu} - i \epsilon^{\alpha \beta \mu \nu}
   k_\alpha p_{b,\beta}\Big) \nonumber \\[3pt] & &
 + (p_t \cdot p_b) \Big( k^\mu p_t^\nu + k^\nu p_t^\mu
 - k \cdot p_t g^{\mu\nu} \Big) \nonumber \\[3pt] & &
 + (k \cdot p_t) \Big( (p_b + k)^\mu p_t^\nu +(p_b + k)^\nu p_t^\mu
 + (p_b + k) \cdot p_t g^{\mu\nu} \Big) \nonumber \\ & &
 - (k \cdot p_b) \Big( 2 p_t^\mu p_t^\nu - m_t^2 g^{\mu\nu} \Big)
 + (k \cdot p_t) \Big( 2 p_b^\mu p_b^\nu - m_b^2 g^{\mu\nu} \Big)
   \nonumber \\[3pt] & &
 - i \Big( \epsilon^{\alpha \beta \mu \nu} k \cdot p_t
 + \epsilon^{\alpha \beta \gamma \mu} k^\nu p_{t,\gamma}
 - \epsilon^{\alpha \beta \gamma \nu} k^\mu p_{t,\gamma} \Big)
   p_{b,\alpha} p_{t,\beta} \nonumber \\ & &
 + i \Big( \epsilon^{\alpha \beta \mu \nu} m_t^2
 + \epsilon^{\alpha \beta \gamma \mu} p_t^\nu p_{t,\gamma}
 - \epsilon^{\alpha \beta \gamma \nu} p_t^\mu p_{t,\gamma}\Big)
   k_\alpha p_{b,\beta} \Bigg\} \nonumber \\ & &
 + \frac{1}{2} B^{\mu\nu} \Delta_{\rm SPF},
 \end{eqnarray}
 where
 \begin{equation}
   B^{\mu\nu} = 2 \left(
   p_t^\mu p_b^\nu +p_t^\nu p_b^\mu -p_b \cdot p_t g^{\mu\nu}
 + i \epsilon^{\mu \nu \alpha \beta} p_{b,\alpha} p_{t,\beta} \right)
 \end{equation}
 is the Born term contribution and the soft photon factor
 $ \Delta_{\rm SPF} $ is given by
 \begin{eqnarray}
   \lefteqn{ \Delta_{\rm SPF} \ = \ - e^2 \Bigg(
   \frac{Q_t^2 m_t^2}{(p_t \cdot k)^2}
 + \frac{Q_b^2 m_b^2}{(p_b \cdot k)^2}
 + \frac{Q_W^2 m_W^2}{(q \cdot k)^2} } \nonumber \\ & &
 - \frac{2 Q_t Q_b \, p_t \cdot p_b}{(p_t \cdot k)(p_b \cdot k)}
 - \frac{2 Q_t Q_W \, p_t \cdot q}{(p_t \cdot k)(q \cdot k)}
 + \frac{2 Q_b Q_W \, p_b \cdot q}{(p_b \cdot k)(q \cdot k)} \Bigg).
 \end{eqnarray}
 $ Q_t = 2/3 $, $ Q_b = -1/3 $ and $ Q_W = 1 $ are the electric charges of the
 top quark, the bottom quark and the $ W $-boson, resp., in units of the
 elementary charge $ e $. The momenta of the top quark, the bottom quark, the
 gauge boson and the photon are denoted by $ p_t $, $ p_b $, $ q $ and $ k $,
 respectively. From momentum conservation one has $ p_t = p_b + q + k $.
 For convenience and generality we have kept $ m_b \ne 0 $ in
 Eq.(\ref{hadrontensor}). It is noteworthy that by setting $ Q_t = Q_b = 1 $,
 $ Q_W = 0 $, $ e^2 = g_s^2 $, and by multiplying by the colour factor
 $ N_C C_F = 4 $ one recovers the QCD tree graph contribution treated e.g.
 in \cite{FGKM2}.

 The transverse-plus, transverse-minus and longitudinal components of the
 hadron tensor can be obtained with the help of the projectors
 $ (\IP^{\mu\nu}_{U+L} - \IP^{\mu\nu}_L + \IP^{\mu\nu}_F)/2 $,
 $ (\IP^{\mu\nu}_{U+L} - \IP^{\mu\nu}_L - \IP^{\mu\nu}_F)/2 $ and
 $ \IP^{\mu\nu}_L $ \cite{FGKM1,FGKM2}, resp., where 
 \begin{equation} \label{projtot} 
   \IP^{\mu \nu}_{U+L} = - g^{\mu\nu} + \frac{q^{\mu}q^{\nu}}{m_W^2} \, ,
 \end{equation}
 \vspace{-5truemm}
 \begin{equation}
   \IP^{\mu \nu}_L = \frac{m_W^2}{m_t^2} \frac{1}{|\vec q\,|^2}
   \Big(p_t^\mu - \frac{p_t\cdot q}{m_W^2} q^{\mu} \Big)
   \Big(p_t^\nu - \frac{p_t\cdot q}{m_W^2} q^{\nu} \Big)
 \end{equation}
 and
 \begin{equation}
   \IP^{\mu\nu}_F = -\frac{1}{m_t} \frac{1}{|\vec q\,|}
   i \epsilon^{\mu \nu \alpha \beta} p_{t,\alpha} q_\beta.
 \end{equation}
 The three components of the hadron tensor are then integrated over two-%
 dimensional phase space. The infrared and collinear singularities are
 regularized with the help of a (small) photon and bottom quark mass,
 respectively. The logarithms of the auxiliary photon and bottom quark masses
 can be seen to cancel against the corresponding logarithms from the one-loop
 contributions in all three helicity components. Details of the calculation
 will be published elsewhere \cite{DGKM}.

 We use the so-called $ G_F $--renormalization scheme for the electroweak
 corrections where $ G_F $, $ M_W $ and $ M_Z $ are used as input 
 parameters~\cite{c14,EMMS91}. The $ G_F $--scheme is the appropiate
 renormalization scheme for processes with mass scales which are much larger
 than $ M_W $ as in the present case. The radiative corrections are
 substantially larger in the so-called $ {\alpha} $--scheme  where $ \alpha $,
 $ G_F $ and $ M_Z $ are used as input parameters~\cite{c14,EMMS91}. In our
 numerical results we shall also present $ {\alpha} $--scheme results along
 with the numerical $ G_F $--scheme results.
 
 Before we present our numerical results on the electroweak corrections we
 briefly discuss the finite width corrections to the Born term rates.
 The finite width corrections are obtained by replacing the $ q^2 $--integration
 over the $ \delta $--function $ \delta(q^2 - m_W^2) $ by an integration over
 the Breit-Wigner resonance curve where the integration is done within the phase
 space limits $ 0 < q^2 < m_t^2 $. The necessary replacement is given by 
 \begin{eqnarray}
   \lefteqn{ \int^{m_t^2}_{0} dq^2
   \delta(q^2 - m_W^2) \,\, \rightarrow} \nonumber \\ & &
   \int^{m_t^2}_{0} dq^2 \frac{m_W \Gamma_W}{\pi}
   \frac{1}{(q^2-m_W^2)^2 + m_W^2 \Gamma_W^2}
 \end{eqnarray}
 where $ \Gamma_W $ is the width of the $ W $-boson
 \hbox{($ \Gamma_W = 2.12 \GeV $)}.

 We are now in the position to present our numerical results. Including the
 QCD one-loop corrections $ \Delta \Gamma_i({\rm QCD}) $ taken from
 \cite{FGKM1}, the electroweak one-loop corrections $ \Delta\Gamma_i({\rm EW})$%
 \footnote{ Since our numerical results are normalized to the Born term rate
 the (small) renormalization of the KM matrix element $ V_{tb} $ \cite{bruecher}
 does not affect our numerical results. We set $ m_H = 115 \GeV $.
 Our results are only very weakly dependent on the Higgs mass.}
 and the finite width corrections
 $ \Delta \Gamma_i({\rm FW}) = \Gamma_i({\rm FW}) -\Gamma_i({\rm zero\ width}) $
 calculated in this paper, and the $ m_b \neq 0 $ corrections to the partial
 Born term rates $ \Delta \Gamma_i(m_b \neq 0) $ \cite{FGKM1} we write
 \begin{eqnarray}
   \Gamma_i & = & \Gamma_i({\rm Born})
 + \Delta \Gamma_i({\rm QCD})
 + \Delta \Gamma_i({\rm EW}) \nonumber \\ & + &
   \Delta \Gamma_i({\rm FW})
 + \Delta \Gamma_i(m_b\neq 0).
 \end{eqnarray}
 It is convenient to normalize the partial rates to the total Born term rate
 $ \Gamma_0 $. The normalized partial rates will be denoted by a hat. Thus we
 write $ \hat{\Gamma}_i = \Gamma_i/\Gamma_0 $ $ (i = +,-,L) $.
 For the transverse-minus and longitudinal rates we factor out the normalized
 partial Born rates $ \hat{\Gamma}_i $ and write ($ i = -,L $)
 \begin{eqnarray}
   \hat{\Gamma}_i & = &
   \hat{\Gamma}_i({\rm Born})(1 + \delta_i({\rm QCD})
 + \delta_i({\rm EW}) \nonumber \\ & + & 
   \delta_i({\rm FW}) + \delta\Gamma_i(m_b\neq 0))
 \end{eqnarray}
 where $ \delta_i = \Gamma_0 \, \Delta \Gamma_i/\Gamma_i({\rm Born}) $.
 Writing the result in this way helps to quickly assess the percentage
 changes brought about by the various corrections.  

 Numerically one has
 \begin{eqnarray}
   \label{rateminus}
   \hat{\Gamma}_{-} & = &
   0.297 (1 - 0.0656 ({\rm QCD}) + 0.0206({\rm EW}) \nonumber \\ & - &
   0.0197 ({\rm FW}) - 0.00172(m_b\neq 0)) \\ & = &
   0.297 (1 - 0.0664)
 \end{eqnarray}
 and
 \begin{eqnarray}
   \label{ratel}
   \hat{\Gamma}_{L} & = &
   0.703 (1 - 0.0951 ({\rm QCD}) + 0.0132 ({\rm EW}) \nonumber \\ & - &
   0.0138 ({\rm FW}) - 0.00357 (m_b\neq 0)) \\ & = & 
   0.703 (1 - 0.0993)
 \end{eqnarray}
  
 \noindent It is quite remarkable that the electroweak corrections tend to
 cancel the finite width corrections in both cases.

 In the case of the transverse-plus rate the partial Born term rate cannot be
 factored out because of the fact that $ \Gamma_{+}(Born) $ is zero. In this
 case we present our numerical result in the form
 \begin{equation}
   \hat{\Gamma}_{+} = \Delta \hat{\Gamma}_{+} ({\rm QCD})
 + \Delta \hat{\Gamma}_{+}({\rm EW})
 + \Delta \hat{\Gamma}_{+}(m_b \neq 0).
 \end{equation}
 \begin{eqnarray}
   \label{rateplus}
   \hat{\Gamma}_{+} & = &
   0.000927 ({\rm QCD}) + 0.0000745({\rm EW}) \nonumber \\ & + &
   0.000358 (m_b\neq 0) = 0.00136.
 \end{eqnarray}
 \noindent Note that the finite width correction is zero in this case.
 Numerically the correction to $ \hat{\Gamma}_{+} $ occurs only at the pro mille
 level. It is save to say that, if top quark decays reveal a violation of the SM
 $ (V-A) $ current structure that exceeds the $ 1\% $ level, the violations
 must have a non-SM origin. Due to the fact that $ \hat{\Gamma}_{+} $ is so
 small the forward-backward asymmetry $ A_{FB}$ is dominantly determined by
 $ \hat{\Gamma}_{-} $ and $ \hat{\Gamma}_L $. We find $ A_{FB}$ = -0.2270.

 To conclude our numerical discussion we also list our numerical results
 for the electroweak corrections in the $ \alpha $--scheme. In the notation
 of Eqs.(\ref{rateminus},\ref{ratel},\ref{rateplus}) we obtain electroweak
 corrections of $ 0.0545 \mbox{(EW)} $, $ 0.0474 \mbox{(EW)} $ and
 $ 6.88 10^{-5} \mbox{(EW)} $ which are $ \approx 62 \, \% $ larger than the
 corresponding corrections in the $ G_F $--scheme.
 \begin{figure} 
  \noindent
  \psfig{file=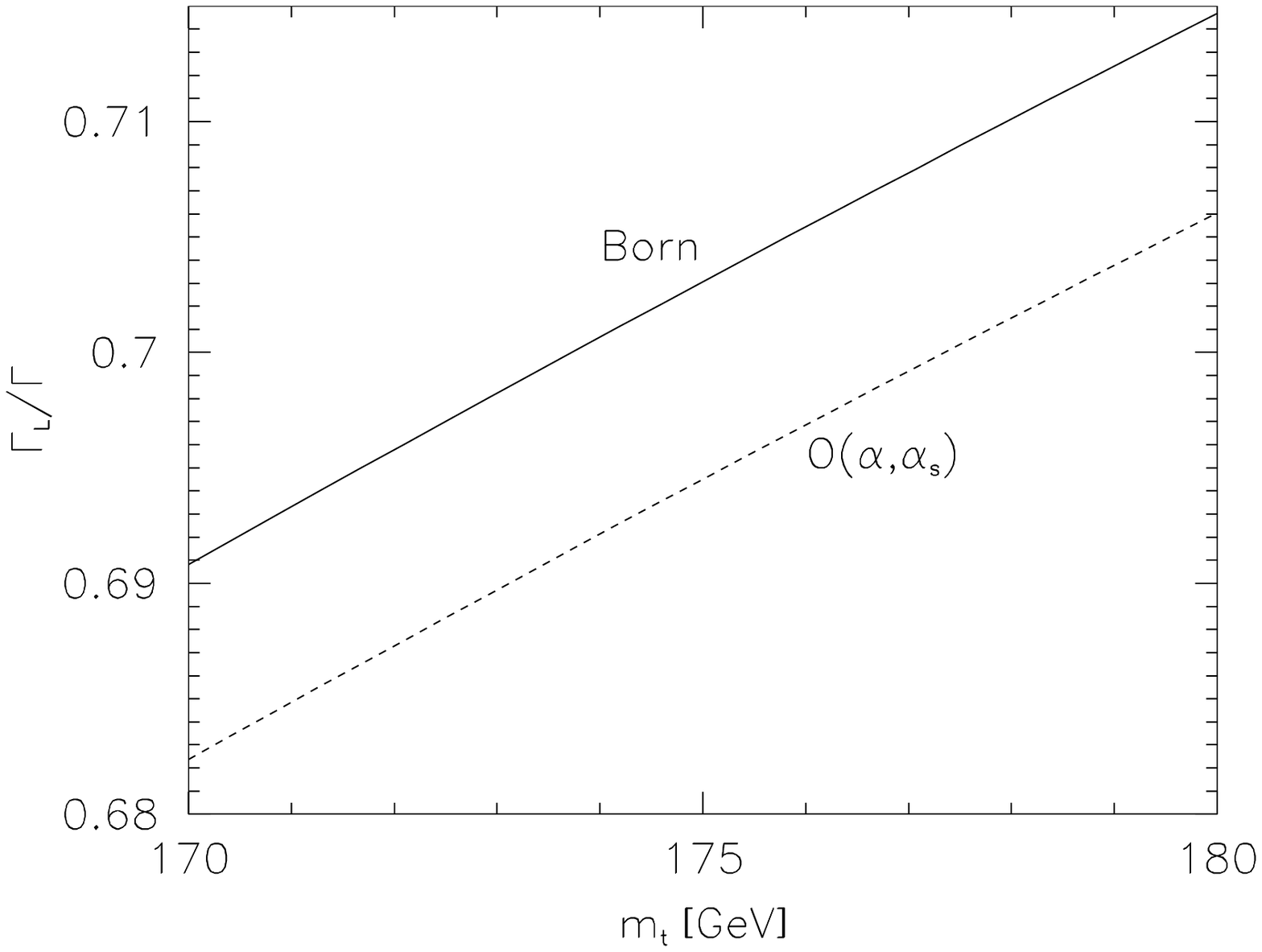,width=8.5cm} \newline
  \psfig{file=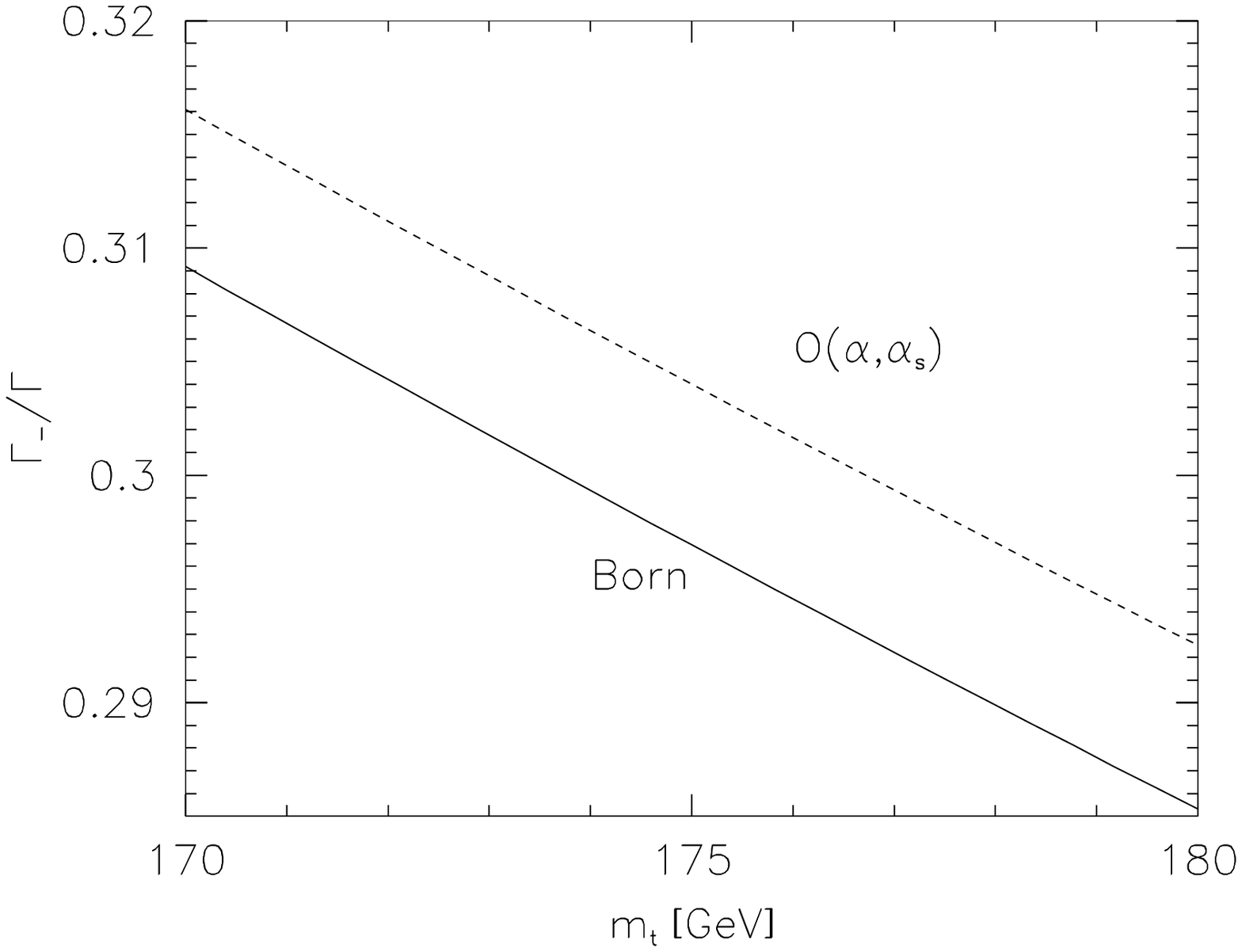,width=8.5cm}
  \caption{\label{plots} Top mass dependences of the ratios
   $ \Gamma_{L}/\Gamma $ and $ \Gamma_{-}/\Gamma $. Full line: Born term.
   Dashed line: Corrections including (QCD),
   electroweak (EW), finite-width (FW) and ($ m_b \ne 0 $) Born term
   corrections.}
 \end{figure}
 In FIG.~2 we show the top mass dependence of the ratio $ \Gamma_L / \Gamma $
 and $ \Gamma_{-} / \Gamma $. The Born term and the corrected curves are
 practically straight line curves. Since the electroweak and the finite width
 effects practically cancel the curves are very close to the QCD corrected
 curves presented in \cite{FGKM1,FGKM2}. The horizontal displacement of the
 respective curves is $ \approx 3.5 $ GeV and $ \approx 3.0 $ GeV for
 $ \Gamma_L / \Gamma $ and $ \Gamma_{-} / \Gamma $. One would thus make the
 corresponding mistakes in the top mass determination from a measurement of
 the angular structure functions if the Born term curves were used instead of
 the corrected curves.
 If we take $ m_t=175 $ GeV as central value a $ 1\% $ relative error on
 the structure function measurement would allow one to determine the top
 quark mass with $ \approx $ 3 GeV and $ \approx $ 1.2 GeV accuracy,
 depending on whether the angular measurement was done on the longitudinal ($L$)
 or transverse-minus ($ - $) mode. Latter result also holds true for the 
 forward-backward asymmetry $ A_{FB} $ since the transverse-plus rate
 is negligible.
 

 In summary, we have calculated the electroweak radiative and finite width
 corrections to the three diagonal structure functions that occur in the
 polar angle distribution of the decay $ t \rightarrow b + W^+ (\rightarrow
 l^+ + \nu_l) $. These will be relevant for a determination of the top quark
 mass. We have not taken into account possible effects coming from the finite
 width of the top quark. These should be smaller than those calculated here
 for the $ W $ width because $ \Gamma_t < \Gamma_W $. Also there can be QED
 and QCD cross-talk between the production and decay processes that could
 spoil the factorization-based angular decay pattern discussed in this paper.
 Both of these effects deserve further studies. 
 
 \vspace{5mm}

 {\bf Acknowledgements:}
 S.~Groote and M.C.~Mauser were supported by the DFG (Germany) through the
 Graduiertenkolleg ``Eichtheorien'' at the University of Mainz.



\begin{thebibliography}{99}
 \vspace*{-8mm}
 \bibitem{cdf} CDF Collaboration, T.~Affolder {\it et al.},\\
  Phys.~Rev.~Lett.\ {\bf 84} (2000) 216
 \bibitem{willenbrock} S.~Willenbrock, hep-ph/0008189
 \bibitem{c14} A.~Denner and T.~Sack,
  Nucl.~Phys.\ {\bf B358} (1991) 46
 \bibitem{c15} J.~Liu and Y.-P.~Yao,
  Int.~J.~Mod.~Phys.\ {\bf A6} (1991) 4925
 \bibitem{c16} A.~Czarnecki,
  Phys.~Lett.\ {\bf B252} (1990) 467
 \bibitem{c17} C.S.~Li, R.J.~Oakes and T.C.~Yuan, \\
  Phys.~Rev.\ {\bf D43} (1991) 3759
 \bibitem{c18} M.~Je\.{z}abek and J.H.~K\"uhn,
  Nucl.~Phys.\ {\bf B314} (1989) 1
 \bibitem{ghinculov} A.~Ghinculov and Y.P.~Yao,\\
  Mod.~Phys.~Lett.\ {\bf A15} (2000) 925  
 \bibitem{EMMS91} G.~Eilam, R.R.~Mendel, R.~Migneron and A.~Soni,\\
  Phys.~Rev.~Lett.\ {\bf 66} (1991) 3105 
 \bibitem{bruecher} A.~Barroso, L.~Br\"ucher and R.~Santos,\\
  Phys.~Rev.\ {\bf D62} (2000) 096003
 \bibitem{CM99} A.~Czarnecki and K.~Melnikov,\\
  Nucl.~Phys.\ {\bf B544} (1999) 520
 \bibitem{CHSS99} K.G.~Chetyrkin, R.~Harlander,
  T.~Seidensticker and M.~Steinhauser,
  Phys.~Rev.\ {\bf D60} (1999) 114015
 \bibitem{FGKLM} M.~Fischer, S.~Groote, J.G.~K\"orner,
  B.~Lampe and M.C.~Mauser,
  Phys.~Lett.\ {\bf B451} (1999) 406
 \bibitem{FGKM1} M.~Fischer, S.~Groote, J.G.~K\"orner and M.C.~Mauser,\\ 
  Phys.~Rev.\ {\bf D63} (2001) 031501
 \bibitem{FGKM2} M.~Fischer, S.~Groote, J.G.~K\"orner and M.C.~Mauser,\\
  Phys.~Rev.\ {\bf D65} (2002) 054036
 \bibitem{pivovarov} A.A.~Penin and A.A.~Pivovarov,
  Nucl.~Phys.\ {\bf B549} (1999) 217; 
  Phys.~Lett.\ {\bf B443} (1998) 264
 \bibitem{jk93} M.~Je\.{z}abek and J.H.~K\"uhn,
  Phys.~Rev.\ {\bf D48} (1993) 1910,
  Erratum {\it ibid}\, {\bf D49} (1994) 4970
 \bibitem{bd02} C.~Bauer and H.S.~Do,
  Comput.\ Phys.\ Commun.\  {\bf 144} (2002) 154
  [arXiv:hep-ph/0102231]
 \bibitem{DGKM} H.S.~Do, S.~Groote,
  J.G.~K\"orner and M.C.~Mauser,\\
  to be published
 \bibitem{Denner} A.~Denner,
  Fortsch.~Phys.\ {\bf 41} (1993) 307
 \end{thebibliography}
 \end{document}